\documentclass[prl,nofootinbib,floats,aps,superscriptaddress,twocolumn]{revtex4}
\usepackage[hypertex]{hyperref}
\usepackage{graphicx} 
\usepackage{amsbsy}   
\usepackage{xspace}

\newcommand{\beq}{\begin{equation}}
\newcommand{\eeq}{\end{equation}}
\newcommand{\beqa}{\begin{eqnarray}}
\newcommand{\eeqa}{\end{eqnarray}}

\def\ra {\rightarrow}
\def\qbq {\overline qq}
\def\qbqqbq {\overline qq \overline qq}
\def\qb {\overline q}
\def\ubu {\overline uu}
\def\dbd {\overline dd}

\def\kbk {\overline KK}
\def\pp {\pi\pi}

\def\sbs {\overline ss}

\begin{document}

\pagestyle{plain}  


\title{\boldmath Reply to ``Comment on `Chiral suppression of scalar 
glueball decay' ''}

\author{Michael Chanowitz}
\affiliation{Ernest Orlando Lawrence Berkeley National Laboratory,
University of California, Berkeley, CA 94720}


\noindent Published in Phys. Rev. Lett. {\bf 98}, 149104 (2007)           

\maketitle

\newpage


In \cite{mc1} I observed that the amplitude for spin zero glueball
decay is proportional to the quark mass, ${\cal M}(G_0 \ra \qbq)
\propto m_q$, to all orders in perturbation theory, so that the ratio
$\Gamma(G_0 \ra \ubu +\dbd)/\Gamma(G_0 \ra \sbs)$ is calculable and
small, even though the individual rates are not perturbatively
calculable because of soft $t$ and $u$ channel quark exchanges. I
noted that if hadronization of $G_0 \ra \qbq$ is an important
mechanism for $G_0 \ra \pp$ and $G_0 \ra \kbk$, then $\Gamma(G_0 \ra
\pp)$ is much smaller than $\Gamma(G_0 \ra \kbk)$, explaining a
previous LQCD result\cite{weingarten} and supporting identification of
$f_0(1710)$ with $G_0$. A more robust consequence, emphasized in
\cite{mc2}, is that mixing of $G_0$ with $\ubu + \dbd$ (and perhaps
also $\sbs$) mesons is suppressed, so that the scalar (and
pseudoscalar) may be the purest glueballs. In both \cite{mc1} and
\cite{mc2} I emphasized the necessity to verify the existence and
consequences of chiral suppression by a reliable nonperturbative
method, which today can only be LQCD.

Chao {\it et al.} agree that $G_0 \ra \qbq$ is chirally suppressed but
propose that $G_0 \ra \qbqqbq$, which is not chirally suppressed, is
the dominant mechanism for $G_0 \ra \pp$. In the preceding
Comment\cite{chm2} and in a previous paper\cite{chm1} they exhibit an
O($\alpha_S$) amplitude for the exclusive process $G_0 \ra \pp$
using light cone wave functions.  Since pQCD for exclusive processes
converges much more slowly than inclusive pQCD\cite{ni-chls}, the
estimate is not quantitatively reliable at the experimentally
interesting scale, $m_G = 1.7$ GeV, where even the applicability of
ordinary inclusive pQCD is marginal. While the $\qbqqbq$ mechanism
might indeed dilute or remove chiral suppression of $G_0 \ra \pp$,
it is not possible to decide, since the magnitude of neither
the $\qbq$ nor $\qbqqbq$ contributions are reliably calculable.

Comparing the amplitudes for ${\cal M}(G_0 \ra \qbq)$ and ${\cal
M}(G_0 \ra \qbqqbq \ra \pp)$ in \cite{mc1} and \cite{chm1,chm2} it
appears that both begin at first order in $\alpha_S$, but this
impression is misleading. It is easy to see that ${\cal M}(G_0 \ra
\qbqqbq \ra \pp)$ vanishes in the chiral limit at O($\alpha_S$) for
on-shell constituent gluons. The $\qbqqbq$ mechanism requires the
quark from one gluon to combine with the antiquark from the other
gluon to form a color singlet pion. But $G_0$ cms (center of mass)
kinematics then requires both quarks to have the same energy fraction,
$x= 2E_q/m_G$ and both antiquarks to have fraction $1-x$, with
$m_{\pi}^2 = x(1-x)m_G^2$.  One of the $q$ or $\qb$ constituents of
each pion is then moving in the opposite direction to the pion in the
$G_0$ cms.  Boosting to an infinite momentum frame, one constituent is
then at $x=1$ and the other at $x=0$, where the wave function
vanishes. In the chiral limit, $m_{\pi}=0$, this is already apparent
in the $G_0$ cms. Since confining dynamics may put the gluons
off-shell of order $\Lambda_{\rm QCD}$, the amplitude does not
actually vanish but is suppressed of order O($\Lambda_{\rm QCD}/m_G)$.

In the revised Comment the authors have responded to this
observation with the added stipulation that the $G_0$ constituent
gluons are maximally off-shell, of order $m_G$. Although this
requirement was not imposed in \cite{chm1}, the result is apparently
unchanged. Certainly one consequence is that $f_g$, the effective
$G_0gg$ coupling, cannot be identified with the corresponding coupling
$f_0$ in \cite{mc1} as is claimed in \cite{chm2,chm1}, but reflects
the off-shell tail of the $G_0$ wave function {\em} or implicitly
contains a factor $\alpha_S$ at the hard scale $m_G$ reflecting 
hard $gg \ra g^*g^*$ scattering to push the gluons maximally
off-shell. Alternatively, hard scattering of $\qbq\qbq$ can align the
quarks suitably with the final state pions, with the amplitude then
explicitly of order O($\alpha_S^2$).

The relative magnitude of the $\qbq$ and $\qbqqbq$ mechanisms for $G_0
\ra \pp$ is not obvious. For the $\qbq$ mechanism we do not know the
magnitude of ${\cal M}(G_0 \ra \qbq)$ because both $\alpha_S(Q)$ and
the running mass $m_q(Q)$ are evaluated at a soft scale,
O($\Lambda_{\rm QCD}$), and thus are not under perturbative
control. In addition we do not know the hadronization rate from $\qbq$
to $\pp$ and $\kbk$ compared to multi-meson final states. On the other
hand, $\Gamma(G_0 \ra \pp)$ via the $\qbqqbq$ mechanism cannot be
reliably estimated and is additionally suppressed by the square of the
coupling, $\alpha_S(Q)^2$, evaluated at the largest scale in the
problem, $Q=m_G$.  It is then important to stress the agreement,
expressed in both \cite{mc1,mc2} and \cite{chm2}, on the most
important point: reliable nonperturbative methods are needed to
determine whether $G_0 \ra \pp$ is chirally suppressed. We eagerly
await LQCD ``data'' and data from BES II to clarify the issue.

Acknowledgments:
This work was supported by the 
Director, Office of Science, Office of High Energy Physics of the 
U.S.\ Department of Energy under contract DE-AC02-05CH11231.


\end{document}